# Simultaneous spectral and spatial modulation for color printing and holography using all-dielectric metasurfaces

Qunshuo Wei[1*], Basudeb Sain[2*], Yongtian Wang[1†], Bernhard Reineke[2], Xiaowei Li[3], Lingling Huang[1‡], Thomas Zentgraf[2§]

1. School of Optics and Photonics, Beijing Institute of Technology, Beijing, 100081, China
2. Department of Physics, University of Paderborn, Warburger Straße 100, 33098 Paderborn, Germany
3. Laser Micro/Nano-Fabrication Laboratory, School of Mechanical Engineering, Beijing Institute of Technology, Beijing 100081, China

**Abstract**：Metasurfaces possess the outstanding ability to tailor phase, amplitude and even spectral responses of light with an unprecedented ultrahigh resolution, thus have attracted significant interests. Here, we propose and experimentally demonstrate a novel meta-device that integrates color printing and computer-generated holograms within a single-layer dielectric metasurface by modulating spectral and spatial responses at subwavelength scale, simultaneously. In our design, such metasurface appears as a microscopic color image under white light illumination, while encrypting two different holographic images that can be projected at the far-field when illuminated with red and green laser beams. We choose amorphous silicon dimers and nanofins as building components and use a modified parallel Gerchberg-Saxton algorithm to obtain multiple sub-holograms with arbitrary spatial shapes for image-indexed arrangements. Such a method can further extend the design freedom of metasurfaces. By exploiting spectral and spatial control at the level of individual pixels, multiple sets of independent information can be introduced into a single-layer device, the additional complexity and enlarged information capacity are promising for novel applications such as information security and anti-counterfeiting.

[*] These authors contributed equally to the work

---

[†] Email: wyt@bit.edu.cn
[‡] Email: huanglingling@bit.edu.cn
[§] Email: thomas.zentgraf@upb.de

Metasurfaces consisting of subwavelength metallic/dielectric antennas can provide a revolutionized way to achieve full control of light with ultrahigh resolution.[1-5] Possessing the advantages of flexibility, simplicity, subwavelength resolution, low absorption loss along with low fabrication cost,[6] they have shown great promise for achieving a wide variety of practical applications, such as beam shaping,[7-9] phase control,[10, 11] achromatic lenses in the visible wavelengths,[12, 13] invisibility cloaking,[14] data storage,[15, 16] optical security and anti-counterfeiting.[17-19] In particular, by spatially encoding interfacial phase jumps at the subwavelength scale, they allow us to reconstruct 2D or 3D holographic images while realizing wide-angle projection and elimination of high-order diffraction.[20-24] Meta-holograms can be designed to reconstruct different images with polarization, angular and wavelength multiplexing by using anisotropic meta-atoms[25, 26] or spatial multiplexing techniques.[27-30]

Apart from the above mentioned spatial modulation, the spectral response offers another degree of freedom for designing metasurfaces. Typically, the geometric shape (especially the anisotropic geometry under different polarization illumination), material property and spatial arrangement affect the spectral response of a metasurface, which is being reflected in its resonance behavior, implying an alteration of the transmission, reflection, absorption, emission and so on. For instance, the coexistence of strong electric and magnetic resonances within meta-atoms can be utilized for the design of Huygens' metasurfaces with almost uniform transmittance.[31-33] In addition, sharp Fano resonances which provide strong field confinement, may offer an advantage in designing low-threshold dielectric metasurface lasers.[34, 35] Differently, a planar chiral metasurface can exhibit giant circular dichroism and asymmetric transmission by its different spectral responses for left-handedness and right-handedness circularly polarized (CP) lights.[36, 37] In this context, most of the existing spectral modulations correspond to the periodic arrangement of meta-atoms. The spectral responses are accomplished by the resonance mechanisms and their dependence on structural parameters, which need to be decided delicately. Among the existing applications of spectral modulation, generating color printing based on spectral responses of metasurfaces features great potentials.

Color printing based on metasurfaces has opened a new route of producing color images with resolution far beyond the limit of current display technologies. Significant advancements have been accomplished including the realization of color displays at the optical diffraction limit,[38-40] polarization encoded color image,[41] polarization multiplexing color printing[42, 43] and also dynamic multicolor printing.[44] However, currently reported color printing metasurfaces are unable to reconstruct holographic images as they do not encode phase distribution or depth information. Similarly, meta-holograms are not designed to control the spectral response of light, which generally appears as random or featureless pattern under incoherent illumination. For further improving the information capacity and enlarge the design freedom of metasurfaces, it is desirable to modulate simultaneously both the spatial and spectral response within a single-layer metasurface.

Here, we propose and experimentally demonstrate a novel type of meta-device that integrates color printing and holographic wavelength multiplexing within a single-layer all-dielectric metasurface by modulating the spatial phase information and the spectral responses simultaneously. We choose amorphous silicon dimers and nanofins as meta-atoms to build up such metasurface whose spectral responses are optimized to obtain the desired structural color in color printing mode. To simplify the simultaneous spatial modulation, we utilize the Pancharatnam-Berry (PB) phase principle, which uses the orientation angle of each anisotropic meta-atom to manipulate the phase distribution pixel by pixel in the holographic mode. In order to encrypt multiple holographic information into a color printing, we have developed a modified parallel Gerchberg-Saxton algorithm to achieve the region division of the "color-printing indexed" arrangements and obtain the different phase profiles for wavelength multiplexing independently. Such an algorithm is the key to achieve the decoupling of spectral and spatial modulation for color printing mode and holographic mode. By using such a simpler design, we realize the microscopic color printing for the direct imaging of the metasurface and the reconstruction of two holographic images with high transmission efficiencies in the far-field. This method is highly promising to be employed in various practical applications such as data storage, information security, authentication, steganography, and anti-counterfeiting.

**Results**

**Design principle.**

**Figure 1** schematically illustrates the dual working modes, incoherent color printing, and far-field holography, of the designed metasurface. It appears as a microscopic color image in plain view, for example, an earth map, while encrypting multiple holographic projections that can be imaged in the far-field (Fraunhofer regime) under different illumination wavelengths, as holographic multiplexing. The reconstructed holographic images carry entirely different information compared to the color printing, such as 'red blossoms' and 'green leaves' under red and green laser illumination, respectively. To achieve such dual functionality, we develop a simple method to independently separate the spatial and spectral information by using only a single-layer metasurface.

We design "color pixels" that possess both different spectral and distinct spatial phase/amplitude response. The unit-cells of the metasurface need to satisfy three requirements. First, the spectral responses of the meta-atoms have to produce distinct colors. Second, the amplitude responses have to be mutually exclusive at the wavelengths of interest of the desired holographic multiplexing mode to provide adequate efficiencies and avoid crosstalk (color filters). Third, the phase modulation for each pixel needs to act independently to allow a negligible effect on the spectral responses. To fulfill these requirements, we choose anisotropic nanostructures together with circularly polarized illumination to allow a Pancharatnam–Berry (PB) phase modulation while ensuring identical spectral responses. Such a method depends on the polarization conversion history, which has been demonstrated to occur for circularly polarized light when converted to its opposite helicity. Utilizing this phase modulation principle, the acquired phase only depends on the azimuthal angles of the meta-atoms, without affecting the spectral response.

As shown in **Figure 2(a)**, we choose amorphous silicon dimers and nanofins as building blocks to construct the desired metasurface on a silica substrate. The dimers are selected to provide extra design freedoms against the constraint of using a high refractive index of silicon in the visible range and the fabrication accuracy related to the size of the antenna. By varying the height ($H$), length ($L$), width ($W$) of the chosen unit elements and the gap between them, we select the most suitable "amplitude filters" in green and red wavelength ranges for cross circularly polarized light. Each unit cell has a height of 300 nm with a period of 300 nm in both x- and y-directions. The other geometric parameters for the dimers are chosen as length $L_1$ = 90 nm and width $W_1$ = 50 nm with a gap size of 80 nm, while the nanofins have a length $L_2$ = 125 nm and a width $W_2$ = 90 nm. The simulated cross circularly polarized transmittance spectral responses of these two types of optimized meta-atoms by using rigorous coupled-wave analysis (RCWA) method are illustrated in **Figure 2(b)**. Both the spectra show a reasonably good wavelength selectivity with a relatively high cross-polarized transmittance of about 20% and 50% at the desired wavelengths (540 nm for dimers and 645 nm for nanofins, respectively). Note that the intrinsic loss of silicon at the visible range restricts the further improvement of the transmission efficiency through optimization. The crosstalk between the two wavelengths is lower than 5%. To quantitatively evaluate the obtained structural colors, the cross-polarized transmittance spectral responses are converted into the CIE 1931 color space chromaticity diagram. The corresponding colors of dimers and nanofins are marked in the CIE chromaticity chart in **Figure 2(c)**. The entirely different spectral responses of dimers and nanofins enable us to select distinctive colors to reproduce the color printing under white light illumination.

To further confirm our initial assumption of the independent phase modulation based on the PB phase principle of the building blocks, we carried out numerical calculations with the finite-difference-time-domain (FDTD) method. **Figure 2(d)** shows the phase change $\Phi$ and the cross-polarized transmittance of the dimers and nanofins with respect to their orientation angle $\varphi$ at 540 nm and 645 nm. The simulation shows that the orientation-controlled phase covers a 0 to $2\pi$ range, while the transmission efficiency remains uniform for all rotation angles. Such geometric phase obeys the relation of $\Phi=2\sigma\varphi$, which is solely determined by the orientation angle $\varphi$ and incident helicity $\sigma$ of the light. Hence, these two types of meta-atoms can fulfill the requirements for generating simultaneous color printing and holography.

**Calculation of the computer-generated hologram.**

In our design, the spectral responses of amorphous silicon dimers and nanofins correspond to green and red color, respectively. Different types of meta-atoms need to be allocated to the individual areas of different colors. The arrangements of different meta-atoms can additionally encode a chosen color pattern while the orientation angles record the phase information of each hologram. Meanwhile, such spatial phase distribution will not affect the recorded color printing images, which is ensured by the circularly polarized light illumination. However, although holograms can still reconstruct the same corresponding images after scratch or breakage due to redundancy, the loss of information caused by the reduction of pixel number will still deteriorate the quality of the reconstructed images. Therefore, we cannot directly superimpose different sub-holograms with the index of color-pattern to achieve the ideal effect. To solve this problem, we applied a modified parallel iterative Gerchberg-Saxton algorithm that can obtain multiple holograms with arbitrary pixel arrangements (**Figure 3**). With such an algorithm, we obtain an optimized phase-only hologram which enables to reconstruct different target objects with good wavelength selectivity.

The design flow chart for the algorithm is as follows: First, we choose the patterns of different color channels for color printing by exploiting the freedom of dividing the metasurface into regions of arbitrary shapes. In our method, the area assigned to each hologram is indexed by the color printing pattern. Second, we construct parallel iterative loops between the hologram planes and the object planes of different wavelengths, independently via the Fourier transform propagating function. Taking into account the "color-printing indexed" arrangements of different regions, our algorithm only retains the phase distributions within the irregular shapes, while setting the amplitude and phase to be zero at all the other regions. Meanwhile, we introduced an amplitude replacement feedback function at the reconstruction planes to accelerate the convergence of the method, which can be expressed as $T_n = T + |T - T_n'|\kappa$, where $T$ is the target object, $T_n$' is the amplitude calculated from the $n^{th}$ iteration loop, and $T_n$ is the result of feedback function which used to replace $T_n$'. By appropriately choosing the $\kappa$ value in this formula, the feedback operation can increase the convergence speed more efficiently.[24] Finally, we combine the color channels together to form the final integrated hologram. Because each sub-hologram is optimized to the shape of the corresponding color pattern of the color printing, the crosstalk can be avoided to the largest extent. Moreover, the orientation of the meta-atoms does not affect their spectral responses, their exclusive responses at the two working wavelengths further ensure the reconstruction quality of each holographic image. Based on the Fermat principle, mathematically, the relation between different reconstruction distances can be simplified as $\lambda_1 z_1 = \lambda_2 z_2$, under paraxial approximation for different incident wavelengths.[45] In the case of integrating different holographic images to form color holography, we appropriately introduced pre-compensation processing according to the above formula by considering wavelength-dependent magnification to form the uniform size.

**Experimental results.**

We characterize the performance of our dual-functional meta-device by using an experimental set-up, shown in **Figure 4(a)** (see Materials and methods for detailed description). Two differently designed all-dielectric silicon meta-devices were fabricated on top of a glass substrate by using a plasma etching process, following an electron beam lithography for patterning (see Materials and methods). The size of the metasurfaces was 200 μm × 200 μm, containing 666 × 666 pixels, composed of both amorphous silicon dimers and nanofins with different orientation angles. Three exemplary scanning electron microscopy (SEM) images of the samples are shown in **Figure 4(b)**. From these images, different regions and a mixture of nanofins and dimers are observed, which correspond to the color patterns, while the rotation angle within each pixel provides the degree of freedom for the holographic operation mode.

**Figure 5(a)** and **(b)** demonstrate numerical and experimental observations of a clear bicolor pattern of the "earth map", which is based on the spectral response of the metasurface when illuminated the first sample with a broadband white light continuum from a tungsten light source. The dimers that were arranged within the aquatic region of the earth map appear with a dark green color in the experiment, while the nanofins that were arranged in the land region appear with an orange color. The color difference between the designed pattern and the fabricated sample (experimental observation) are mainly due to the deviation of material property and fabrication deviations from the simulations (details can be found in the Supplementary Material). However, illuminating simultaneously by using a green and a red laser with circularly polarized light, this metasurface yields the target holographic images of "red blossoms" and "green leaves" with high resolution and matched magnifications as shown in **Figure 5(c)** and **(d),** respectively.

Furthermore, we investigated the spectral response of the meta-device at different wavelengths ranging from 500 nm to 690 nm by using a supercontinuum laser Fianium WL-SC-400-2 (**Figure 5(e)**). The green leaves image without the obvious crosstalk of green blossoms can be observed for the wavelength range between 510 and 540 nm, and the red blossoms without the disturbance of red leaves can be captured at the wavelengths, ranging from 640 to 670 nm. The peak of the spectral response for "leaves" slightly deviates from the designed value of 540 nm as to maintaining the fabrication precision of the dimer structures was more challenging than the nanofin structure. Illuminating by the wavelengths ranging from 550 to 630 nm, both images are observed to be reconstructed simultaneously. This is due to the fact that although the two types of meta-atoms show reasonably good wavelength selectivity, they still have a non-zero transmittance at the certain spectral range. At these wavelengths, the intensity difference between the two reconstructed images of the "leaves" and the "blossoms" is not significantly large, so the undesired images appear, resulting in the crosstalk between the two holographic channels and low contrast. A comprehensive analysis can be found in the Supplementary Material.

In addition, the second sample demonstrate a QR-code as color printing under broadband white light illumination, while, two other reconstructed holographic images, "Color Printing" and "Metasurface Holography", are observed under two different illumination wavelengths in the *k*-space. The crosstalk-free images "Metasurface Holography" and "Color Printing" with satisfactory quality can be captured at the wavelengths, ranging from 520 to 540 nm and 630 to 650 nm, respectively, which show a reasonably good wavelength selectivity (see Supplementary Material). Noticeably, both experiments demonstrate that the holographic reconstructed images cannot be inferred from the color printing patterns, implying, the spectral response is totally independent of the spatial response in such a simple design. This feature enables us to realize an optical security device by hiding secret information and many other applications.

**Discussion**

Some recent works have demonstrated the possibility of spatial and spectral light modulation with metasurfaces. For example, by utilizing a multi-layer design like a phase plate together with amplitude filters, composed of dielectric pillars, the structural color and holography can be achieved.[46] However, such design suffers from the complexity of nanofabrication based on direct laser writing, which has a very low fabrication efficiency due to pixel by pixel carving compared to electron beam lithography. Furthermore, the spectral and spatial modulation properties are not totally separated. Another work demonstrated single holography together with binary colorprint in the reflection-type scheme, which shows insufficient controllability.[47] On the other hand, by utilizing the orthogonal polarization selectivity of cross antennas, in situ anisotropic thermo-plasmonic laser printing for color printing and far-field holography has been demonstrated,[48] but also suffers from low fabrication efficiency. While our method uses a single layer metasurface and successfully decouples the two working modes, which shows advantages in independent spectral and spatial optical modulation as well as for high information capacity applications, strong controllability, and fast processing.

Furthermore, our developed method can further be extended to more versatile color patterns and increased numbers of holographic channels in wavelength multiplexing holography. Due to the limitation of the refractive index (absorption at shorter wavelengths) and the required high fabrication precision of amorphous silicon, one can choose other materials such as titanium dioxide ($TiO_2$) and silicon nitride ($Si_3N_4$), which have lower absorption in the visible wavelengths range. For example, by carefully designing the geometric parameters of different nanostructures made by $TiO_2$, the RGB tricolor pattern can be achieved, and such a metasurface can simultaneously record different holographic information at those three different design wavelengths. Moreover, our metasurfaces can further be extended to achieve multifunctionality for arbitrarily shaping the wavefronts while independently controlling the spectral information through the individual pixel modulation.

In summary, by exploring the spatial and spectral design freedom, we successfully demonstrate single-layer all-dielectric silicon metasurfaces that function at dual working modes, that is, color printing and meta-holography. The metasurfaces consist of two types of meta-atoms made by amorphous silicon, each of them acts as a color filter under white light and provides a color channel for a specific wavelength to independently manipulate phase distributions by utilizing their orientations angles. Both the phase and spectral responses can be defined at a subwavelength scale simultaneously and independently. We developed a modified parallel iterative Gerchberg-Saxton algorithm, which obtains holograms for arbitrary shapes to adapt "color-printing" indexed pattern. Such an algorithm is the key to the wavelength multiplexing holograms by utilizing the color filter property (wavelength selectivity) of the two designed meta-atoms. Owing to the large information capacity and the novel dual-mode design, our meta-device may open promising applications in optical security and encryption, anti-counterfeiting, high-resolution holographic data storage, optical information processing, and many other fields.

## Methods

**Fabrication of the meta-devices.**

The all-dielectric silicon metasurfaces were fabricated on a glass substrate following the processes of deposition, patterning, lift-off, and etching. First, through plasma-enhanced chemical vapor deposition (PECVD), we prepared a 300-nm-thick amorphous silicon (a-Si) film. Following this, a poly-methylmethacrylate resist layer was spin-coated onto the a-Si film and baked on a hot plate at 170°C for 2 min to remove the solvent. Next, the desired structures were patterned by using standard electron beam lithography. Subsequently, the sample was developed in 1:3 MIBK:IPA solution and then washed with IPA before being coated with a 20-nm-thick chromium layer by electron beam evaporation. Afterward, a lift-off process in hot acetone was performed. Finally, by using inductively coupled plasma reactive ion etching (ICP-RIE), the desired structures were transferred from chromium to silicon.

**Design and numerical simulations.**

In order to obtain the spectral responses that satisfy all the three requirements, we carried out a 2D parameter optimization by using the RCWA method. The corresponding refractive index of amorphous silicon was experimentally measured by ellipsometry. The length (*L*) and width (*W*) of the dimers and nanofins were swept in the range of 50 to 200 nm and 50 to 125 nm, respectively, while maintaining the height (*H*) as 300 nm and the period as 300 nm to eliminate undesired orders of diffraction. Besides, the gap size of the dimers was also swept from 50 to 100 nm by considering the fabrication precision. To further confirm the initial assumption of the independent phase modulation based on PB phase principle of the chosen meta-atoms, we also carried out numerical calculations with FDTD method. For the simulation, the dimers and nanofins were placed onto a glass substrate ($n_{SiO2}$ = 1.46). Periodic boundary conditions were employed in both *x*- and *y*-axis, and the perfectly matched layers were applied in the *z* direction. The wavelength of incident light is fixed at 540 nm and 645 nm for the dimers and nanofins, and the corresponding refractive index of amorphous silicon was experimentally measured by ellipsometry.

**Optical measurement.**

For the optical characterization of the performance of our dual-functional meta-device, we use the setup that is shown in Figure 4(a). Considering the property of PB phase modulation mechanism, a combination of a linear polarizer (LP) and quarter-wave plate (QWP) is positioned in front of and behind the sample to prepare and select the desired circular polarization state for the incident/transmitted light. A magnifying microscope objective (40×/NA=0.6) is positioned behind the sample to capture the images. In color printing mode, dual-color patterns can be directly observed under white light illumination or in any optical microscope. Whereas in the case of the holographic images the reconstruction appears in the *k*-space, we use the objective together with a lens to observe the Fourier plane on a CCD camera. Moreover, the magnifying ratio and numerical aperture of the objective lens are carefully chosen for the purpose of collecting all the diffraction light from the sample and reconstructing the holographic images in the Fourier plane. We use a supercontinuum laser for the broadband evaluation and red and green lasers for single wavelength illumination.

**Acknowledgments**

The authors acknowledge the funding provided by the National Key R&D Program of China (No. 2017YFB1002900) and the European Research Council (ERC) under the European Union's Horizon 2020 research and innovation program (grant agreement No. 724306). We also acknowledge the NSFC-DFG joint program (DFG No. ZE953/11-1, NSFC No. 61861136010). L.H. acknowledge the support from National Natural Science Foundation of China (No. 61775019) program, Beijing Municipal Natural Science Foundation (No. 4172057), Beijing Nova Program (No. Z171100001117047) and Fok Ying-Tong Education Foundation of China (No.161009).

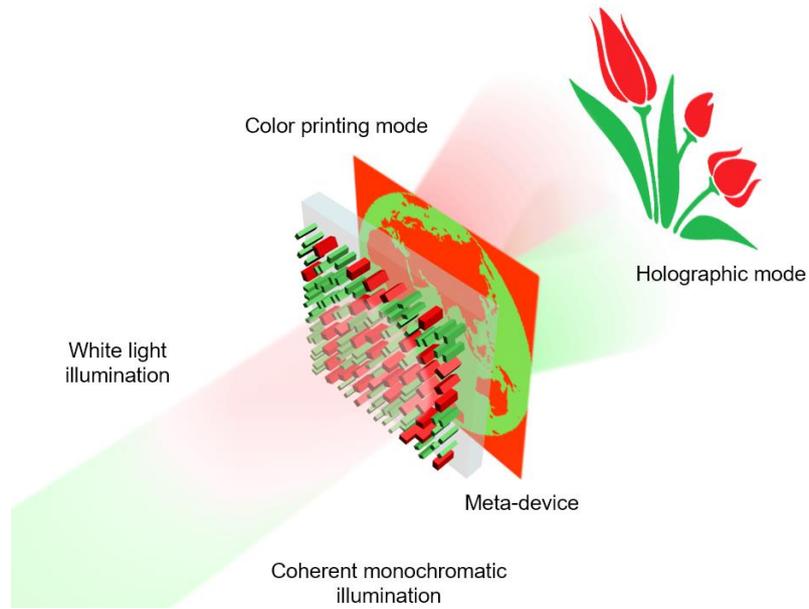

**Figure 1.** Schematic illustration of the all-dielectric metasurface that integrates dual working modes for incoherent color printing and far-field holography by modulating spatial and spectral responses simultaneously. The metasurface is composed of amorphous silicon dimers and nanofins with optimized spectral responses to obtain the desired structural color. When illuminating with different wavelengths, it can reconstruct different encoded holographic images in contributed as a multiplexing hologram.

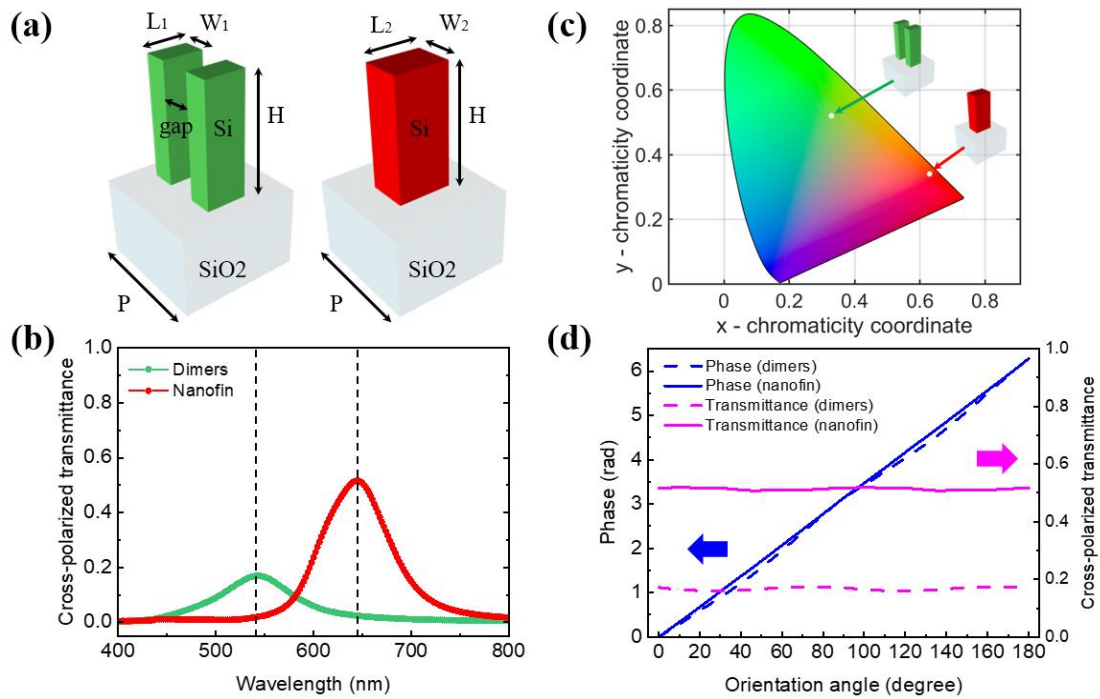

**Figure 2.** (a) Schematic of the two types of meta-atoms on glass substrates. (b) The cross circularly polarized spectral responses of amorphous silicon dimers and nanofins. Both spectrums have a relatively high cross-polarized transmittance of about 20% and 50% at the desired wavelengths (540 nm for dimers and 645 nm for nanofins, marked out by dash lines), and the crosstalk between the two wavelengths are lower than 5%. (c) The calculated structural colors in the CIE 1931 chromaticity diagram from the simulated transmittance spectral responses of dimers and nanofins. (d) The phase change and the cross-polarized transmittance of dimers and nanofins with respect to their orientation angle at 540 and 645 nm.

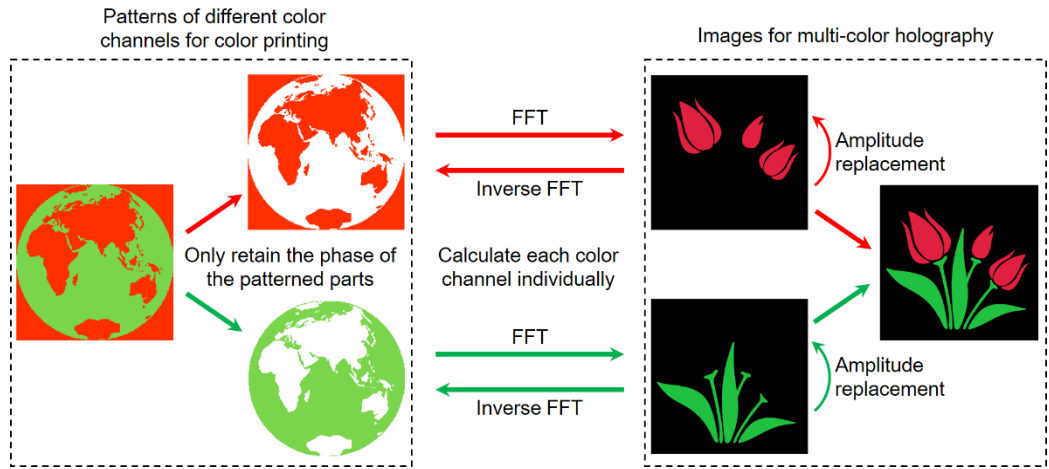

**Figure 3.** The flow chart of the modified parallel Gerchberg-Saxton algorithm which can obtain multiple holograms with arbitrary shapes. "FFT" and "Inverse FFT" represent Fresnel transform and inverse Fresnel transform in the light propagation process.

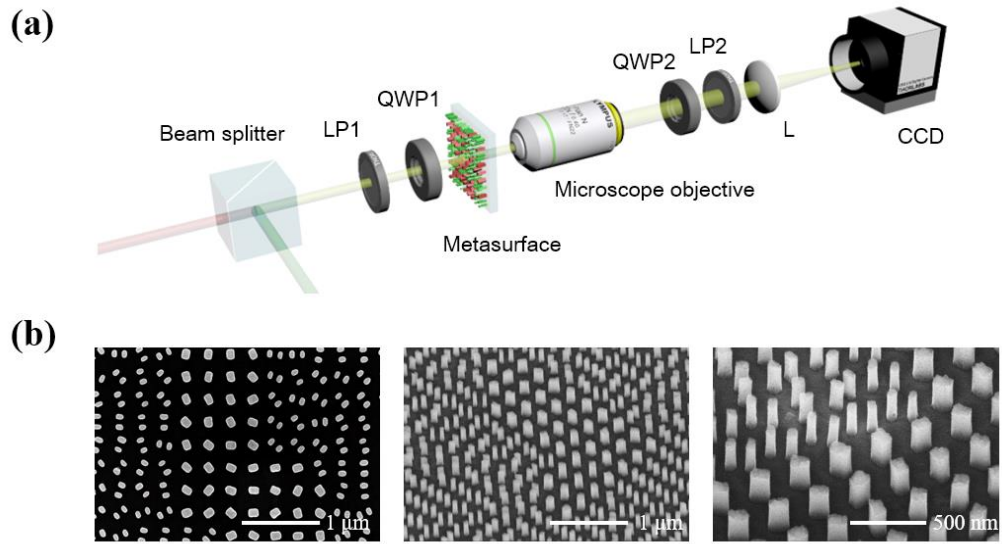

**Figure 4.** (a) The experimental set-up for observation of the color printing patterns and reconstruction of the holographic images. In color printing mode, dual-color patterns can be directly observed through the microscopic arrangement under white light illumination. For the reconstruction of the encoded holographic images in the far-field when illuminating with red and green laser beams, the observation is set to the Fourier plane. (b) The scanning electron microscopy (SEM) images (one plane view and two side views with different magnifications) of fabricated metasurfaces.

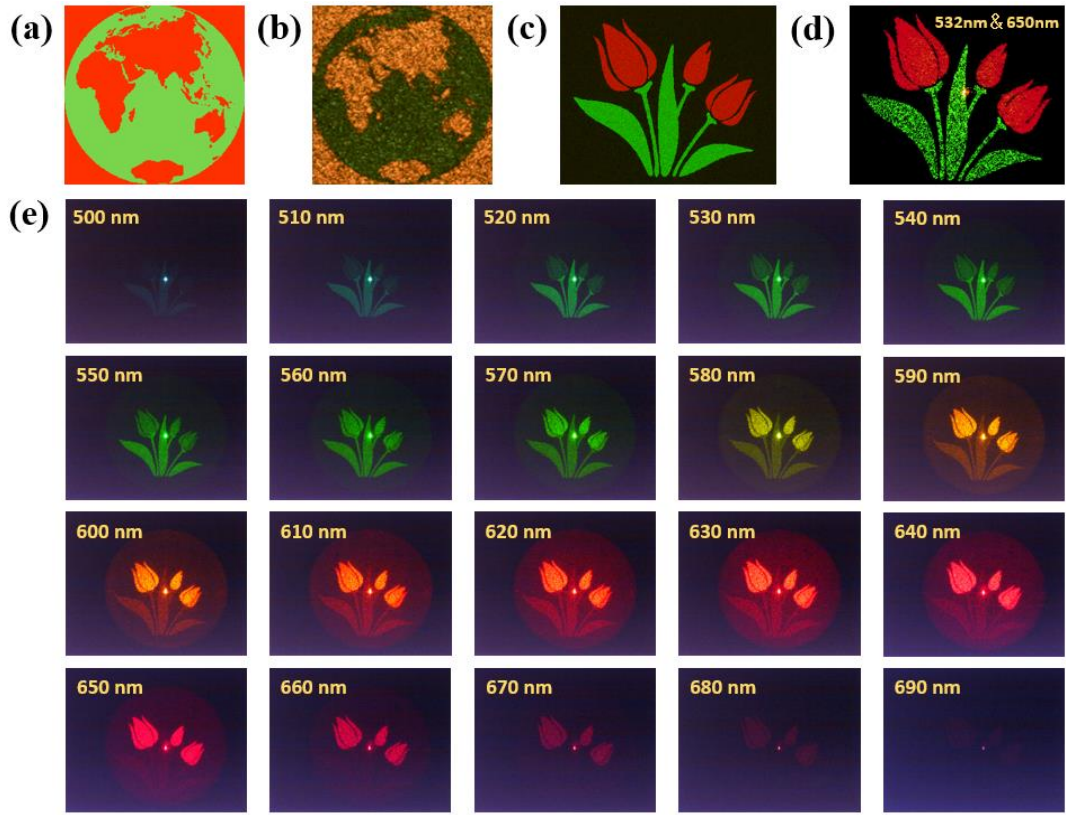

**Figure 5.** The design and the experimental results of our sample that integrates color printing and meta-hologram within a single-layer all-dielectric metasurface. (a) The bicolor pattern of the "earth map" which was used in color printing mode. (b) The experimental result of color printing mode. The pattern of the "earth map" is a microscopic image of the cross-polarized white light. (c) The simulation results of the target images "red blossoms" and "green leaves" for the holographic reconstruction. (d) The experimentally reconstructed holographic images of the "red blossoms" and the "green leaves" when simultaneously illuminating with red (532 nm) and green (650 nm) laser beams. (e) The broadband spectral property of metasurface under holographic mode by using a supercontinuum laser with wavelengths of 10 nm bandwidth, ranging from 500 to 690 nm.